\documentclass[journal]{IEEEtran}

\IEEEoverridecommandlockouts
\usepackage{cite}
\usepackage{amsmath,amssymb,amsfonts}
\usepackage{algorithmic}
\usepackage{graphicx}
\usepackage{textcomp}
\usepackage{xcolor}
\usepackage{caption}
\usepackage{subcaption}
\usepackage{threeparttable}
\usepackage{glossaries}
\usepackage{comment}
\usepackage{booktabs}

\newacronym{ssa}{SSA}{shared situation awareness}
\newacronym{smm}{SMM}{shared mental model}
\newacronym{ai}{AI}{artificial intelligence}
\newacronym{loa}{LOA}{level of abstraction}
\newacronym{EDL}{EDL}{entry, descent, and landing}
\newacronym{xai}{XAI}{explainable AI}
\newacronym{aa}{AA}{Automated/Autonomous Agent}
\newacronym{amp}{AMP}{Automated/Autonomous Mission Planner}
\newacronym{ndm}{NDM}{naturalistic decision making}

\begin{document}
%\title{The Impact of Human Shared Situation Awareness in Unreliable AI-Advised Decision Making}
%\title{The Impact of Human Shared Situation Awareness in Recommender Systems}

\title{Aligning Judgment Using Task Context and Explanations to Improve Human-Recommender System Performance}

\author{Divya~Srivastava,~\IEEEmembership{Student Member,~IEEE,}
        and~Karen M.~Feigh,~\IEEEmembership{Senior Member,~IEEE}% <-this % stops a space
\thanks{D. Srivastava is with the Department
of Mechanical Engineering, Georgia Institute of Technology, Atlanta,
GA, 30332 USA e-mail: (divya.srivastava@gatech.edu).}}


\markboth{Transactions on Human Machine Systems,~Vol.~X, No.~X, xXX~20XX}%
{Shell \MakeLowercase{\textit{et al.}}: Bare Demo of IEEEtran.cls for IEEE Journals}



\maketitle

\begin{abstract}
Recommender systems, while a powerful decision making tool, are often operationalized as black box models, such that their AI algorithms are not accessible or interpretable by human operators. This in turn can cause confusion and frustration for the operator and result in unsatisfactory outcomes. While the field of explainable AI has made remarkable strides in addressing this challenge by focusing on interpreting and explaining the algorithms to human operators, there are remaining gaps in the human's understanding of the recommender system. This paper investigates the relative impact of using context, properties of the decision making task and environment, to align human and AI algorithm understanding of the state of the world, i.e. judgment, to improve joint human-recommender performance as compared to utilizing post-hoc algorithmic explanations.
%how providing task context, properties of the decision making environment, influences the human-recommender system's joint performance compared to providing explanations that are reliant on interpreting the algorithms themselves. 
We conducted an empirical, between-subjects experiment in which participants were asked to work with an automated recommender system to complete a decision making task. We manipulated the method of transparency (shared contextual information to support shared judgment vs algorithmic explanations) and record the human's understanding of the task, the recommender system, and their overall performance. We found that both techniques yielded equivalent agreement on final decisions. However, those who saw task context had less tendency to over-rely on the recommender system and were able to better pinpoint in what conditions the AI erred. Both methods improved participants' confidence in their own decision making, and increased mental demand equally and frustration negligibly. 
These results present an alternative approach to improving  team performance to post-hoc explanations and illustrate the impact of judgment on human cognition in working with recommender systems. 
%underscore the critical role of transparency in AI systems for enhancing team performance, and furthermore, this approach of providing context enables a range of solutions for improving transparency in human-AI collaborations.

%This paper investigates the relative impact of using context to align human and AI algorithm understanding of the state of the world, i.e. judgement, to improve recommender performance as compared to utilizing post-hoc explanations.  An empirical,  between-subjects experiment in which participants were asked to work with an automated recommender system to complete a decision making task.  We manipulated the task transparency method - shared contextual information to support shared judgment and local explanation.  Results indicated that both methods equally improved Final Decision Agreement however, the shared judgment method was found to improve Task Performance and the Resolution of Team Disagreements.  Both methods increased mental demand equally.  
\end{abstract}


\section{Introduction}
\IEEEPARstart{R}{ecommender}
systems offer immense value in guiding human decision-making across critical domains. Traditionally, these systems employed algorithms designed to suggest optimal courses of action to users. 
%However, the field has witnessed a remarkable progression, with sensors and algorithms becoming increasingly sophisticated over time. 
The latest iteration of recommender systems use pre-trained AI models to form the foundation of their automated agents and recommendation systems. These cutting-edge systems leverage advanced artificial intelligence to provide highly informed and tailored guidance, empowering humans to make well-informed choices in low-stakes \cite{low-stakes-ex} and high-stakes \cite{high-stakes-ex} scenarios.

While recommender systems offer immense potential, their increasing sophistication, particularly those powered by AI-trained algorithms, has resulted in a significant challenge – opacity. Unlike their predecessors, these advanced systems are virtually impenetrable to human comprehension \cite{lipton-2017}, utilizing algorithms that are either not available to or not easily interpretable by humans. These systems are often characterized as ``black boxes" \cite{vonEschenbach2021}. Further, this opaqueness is purposefully implemented and valued as it seemingly facilitates faster decision-making \cite{Baker1995BreastCP, parmar2021}. However, this approach can be problematic. In well-defined, simple decision environments or when the human operator is an expert, opaque AI systems may yield satisfactory results \cite{martignon2002fast}. However, in complex scenarios or challenging decision landscapes, these systems are prone to failure \cite{canellas2015}. When the AI's recommendation is unsatisfactory or even catastrophic \cite{NTSB_accidentreport2014, williams2021}, users are left with little to no insight into the reasoning behind the suggested course of action, making it difficult to recognize \cite{parasuraman2010complacency, zsambok1997naturalistic} or rectify errors.  And in many safety critical domains, these random failures represent unacceptable risks which must be managed.

Recognizing these challenges, significant efforts have been made to improve the transparency and interpretability of recommender systems. These include extensive training for human operators\cite{yang2019study}, enhanced visualizations of proposed solutions \cite{becker2001visualizing, erra2011interactive}, ranking candidate solutions for comparison \cite{monroe2019}, and attempting to explain the underlying reasoning behind recommendations \cite{Dazeley_2021}. While these approaches have improved specific types of recommender systems, identifying best practices for effective human-AI interaction in recommender systems remains an open challenge.

Much of the work thus far focuses on ``opening" the black box in order to make the automation explainable or more interpretable by humans. However, predominant XAI work focuses on providing \textit{post-hoc} explanations of algorithmic models, which is may not be useful in real-time decision making. While \textit{post-hoc} XAI has been successful with simple automation in previous generations of technology, autonomous systems without explanatory mechanisms are being utilized in present day anyway, which can be detrimental to current users.

Previous work \cite{srivastava2022, srivastava_lilly2024, kolb2023effects} has shown that supporting user judgment by providing contextual information about the decision environment has a positive effect on the human working with the AI-driven recommender systems and their overall combined performance. This contextual information can be any relevant information outside of the black box AI that may serve to increase understanding of the decision environment. This transparency technique yields several benefits such as improved human judgement, improved team performance with the AI partner, regardless of the accuracy of the partner, and more accurately calibrated trust in the AI partner. Follow-up studies indicate that the provision of this contextual information produces these benefits even when the information is presented at different levels of abstraction \cite{srivastava2024}. Since this transparency technique is effective in its goals and robust to abstraction, we next aim to see if it can produce equivalent performance as the popular transparency technique of providing  explanations.

This paper investigates the effectiveness of aligning team judgement using contextual information with respect to the effectiveness of using explanations to support a human working with black box recommender systems. The driving force behind providing relevant context information about the decision environment is to support the holistic judgement of the human decision maker, whereas explanations provide insights into the decision making process of AI algorithms for specific instances or inputs. There are several limitations of trying to extract explanations of the AI algorithms:

\begin{itemize}
    \item \textbf{Complexity}: AI algorithms, especially deep learning models, can be highly complex with numerous layers and parameters. Extracting explanations from such models can be challenging due to their intricate structure and the difficulty of interpreting the relationships between inputs and outputs.
    \item \textbf{Interpretability}: Explanations often rely on simplified models or techniques to approximate the behavior of the AI algorithm. While these approximations can provide insights, they may not capture the full complexity of the original model, leading to potential discrepancies and limitations in the explanations. \cite{lakkaraju2017}
    \item \textbf{Scalability}: Generating explanations for large datasets or complex models can be computationally expensive and time-consuming. As the size and complexity of the data increase, the process of generating explanations may become impractical or infeasible.
    \item \textbf{Sensitivity to input variations}: Explanations can be sensitive to small changes in inputs, resulting in different explanations for similar instances,  making it challenging to rely solely on  explanations for understanding the overall behavior of the AI algorithm. \cite{lundberg2017}
\end{itemize}

There is substantial ongoing research in the field of \gls{xai} to address these limitations. However, providing contextual information should bypass these issues because the information that is relevant to the task is already being used as input to the AI. Providing contextual information to the human decision maker aligns the human and AI judgement and provides a shared situation awareness that may prove as valuable as post-hoc explanations. We hypothesize that providing contextual information will yield comparable team performance and calibrated trust in the AI partner as using local explanations would. If this is the case, then this model-agnostic method of increasing transparency may prove less time-consuming and costly than current XAI techniques that are model-specific (i.e. aimed at explaining or interpreting the specific algorithms of the AI). This would be beneficial for current and future advanced technology, especially for parties concerned about model confidentiality. 

\section{Background \& Prior Similar Work}

Decision Support Systems (DSS) have been a prevalent application of \gls{ai} across various industries for decades, predating the integration of \gls{ai} methods into their underlying algorithms. 
These systems are designed to assist operators in decision-making tasks with the  goal of enabling better and faster decisions. 
They achieve this by either simplifying the decision space or generating potential solutions, thereby reducing the cognitive burden on human decision-makers. 
A core advantage of decision support systems lies in their ability to encode expert-level knowledge and insights, thus empowering non-expert users with access to specialized information. 
AI-driven recommender systems represent a subset of DSS, wherein an \gls{ai}-based algorithm is used to recommend a specific course of action. 
Specific examples of \gls{ai} algorithms include machine learning \cite{zhang2021}, reinforcement learning \cite{hu2018, ie2019}, and neural networks \cite{gong2016, jing2017}.  

\begin{comment}
The process of decision making has been modeled in several ways \cite{dewey1910, simon1976, Kersten94}, but for the purposes of this paper, we will utilize the language of the well-established OODA loop \cite{hendrick-2009} which models the following cognitive processes:
\begin{itemize}
    \item Observation: the collection of data through sensory perception
    \item Orientation: the analysis and synthesis of data to form one's current mental perspective
    \item Decision: the determination of a course of action based on one's current mental perspective
    \item Action: the physical playing-out of the decision
\end{itemize}
\end{comment}

DSS and the more specific recommender systems leverage standard process models of cognition and decision making which have been modeled in several ways. Most involve cyclic 4 step cognitive processes such as: 
SEAL model (Sense-Explore-Act-Learn) \cite{Rana_2020}, SIDA model (Sense-Interpret-Decide-Act) \cite{haeckel1995}, or the OODA loop (Observe-Orient-Decide-Act) \cite{hendrick-2009}. Others are more complex such as Hollnagel's  COCOM \cite{Hollnagel_2000}, ABC Group's Adaptive toolbox and  contingency models. Most decision making models will converge on a few key elements: observation of the external world, integration of that information into the decision maker’s understanding of present and potential future states, consideration and selection of possible courses of action, followed by implementation of that course of action.

The use of recommender systems in these decision making processes is primarily to speed up the process of choosing a course of action, which is typically done by minimizing  or removing altogether the need for a human to observe or take in information about the state of the world, or to process that information to form some sort of judgement of the current state of that world. 

\subsection{Naturalistic Decision Making (NDM) \& Role of Judgement}

The field of \gls{ndm} focuses on comprehending, modeling, and enhancing human decision-making and cognitive function performance in demanding real-world scenarios.
Several works within \gls{ndm} have demonstrated that context and adequate time allocation for the orientation and judgment of relevant information significantly influences the decision-making process. \cite{harringron-ottenbacher-2009, ben-akiva-2012, vazquez-diz-2019}.

%essentially involves 5 steps: recognizing a problem, gathering (relevant) information, generating possible solutions, making the decision, and finally evaluating the decision, which is only possible if there’s feedback.

In AI-advised decision-making tasks, the allocation of responsibility for each cognitive function in the decision-making process is as follows: \gls{ai} assumes responsibility for the Observing and Orienting phases, and can contribute to the Decision phase. \gls{ai} collects pertinent data (Observe/Sense), and then utilizes this information to generate potential solutions (Orient/Judge/Interpret), or to suggest a decision (Decide). The human operator primarily serves as a safety checkpoint, retaining the responsibilities of making the final decision (Decide) and then implementing it (Act). However, the opaque nature of many systems restricts appropriate understanding of the \gls{ai}'s information observation and orientation processes, thus impeding the development of team shared situation awareness (SSA) due to intentional limitation of human involvement in the processes that build situation awareness.

Furthermore, the absence of a judgment phase for humans in the decision making process often leads to overreliance on system recommendations. This is a common issue observed across various domains \cite{wagner2018}, with both embodied \cite{wagner2016, booth2017} and nonembodied automated systems \cite{lee2004, bussone2015}. 
This can be attributed to a variety of explanations, though Wagner et al conclude that ``people might assume [automation] has knowledge [they] do not possess, viewing [automation]’s actions, at times mistakenly, as a direct reflection of the intentions of the developer, when in fact the [automation] may be malfunctioning or users may be misinterpreting its abilities. Even when presented with evidence of a system’s bad behavior or failure” \cite{wagner2018}. 
Whatever the underlying cause, overreliance can be mitigated through accurate mutual understanding of tasks, roles, and responsibilities between human and AI teammates, such that each can make accurate predictions of each other's decision making capabilities.

\subsection{Model-Specific Contributions to Mitigate Challenges}

The field of \gls{xai} seeks to address challenges posed by the opaque nature of AI systems.
Some approaches aim to provide algorithmic transparency by presenting insights into how the AI processes information. 
Much work in this area focuses on providing explanations for isolated decision points \cite{lakkaraju2017, guidotti2018, Dazeley_2021}.
The most prevalent types of explanation are global explanations and local explanations. Global explanations explain the overall model behavior, e.g. a full set of heuristics or rules used to classify inputs into outputs, whereas local explanations offer a specific explanation for each specific set of decision parameters.

While common, this approach is not without its challenges.
Explanations can be presented in various formats, ranging from textual descriptions to visual representations \cite{guidotti2018}, and there is no consensus on what constitutes a good or effective explanation.
For simpler \gls{ai} algorithms (e.g., decision trees), explanations can be relatively straightforward, often involving the visualization of the decision structure or providing the features that triggered the decision.
In contrast, explanations for more complex algorithms, such as deep neural networks, are challenging to generate as the sheer volume of parameters and nodes used by the algorithm makes extracting a straightforward human-like explanation from them very difficult.
Moreover, the explanation's effectiveness depends on its relevance to the user's existing mental models and its ability to increase the human’s understanding \cite{Dazeley_2021}.
The explanation must be communicated at an appropriate level of abstraction for the individual user.

While successful implementation of this type of algorithmic transparency can provide post-hoc context for outputs, it remains that the decision making process is truncated for the decision maker, as they do not directly \textit{Observe} any information and must work backwards to \textit{Orient} the explanation with the AI's suggestion.

Other work has attempted to assist the human in evaluating the AI's suggestion by providing more interaction with the suggestion, or providing an alternate visualization of the suggestion  \cite{becker2001visualizing, erra2011interactive}.
While these techniques can increase understanding of simple algorithms, they alone may not be sufficient to support the human's judgment in complex decision-making tasks.
All of these approaches make valid contributions to the push to make human-AI decision making teams better, but as stand-alone techniques, they all truncate the decision making process which necessarily creates knowledge gaps for the decision maker, and therefore, skews their judgment.

Recent advancements in  transparency methods have aimed to make complex machine learning models more understandable without uncovering their internal algorithms.
These methods do not rely on the specifics of the machine learning model, and as such, can be applied to any machine learning model after it has been trained.
Two prominent examples of these methods are LIME and SHAP. 
Local Interpretable Model-agnostic Explanations (LIME) \cite{ribeiro2016} works by approximating the behavior of a complex model locally around a specific prediction using a simpler, interpretable model. 
This allows for explanations of individual predictions in terms that humans can understand.
However, this technique faces the same aforementioned challenges associated with explanations.
Shapley Additive Explanations (SHAP) \cite{lundberg2017} are based on game theory concepts and provide post-hoc global explanations of how much influence each input variable has on the target output variable. While insight into the relative importance of input variables within the model can be useful, this technique falls short in conveying what specifically the input variables are. Without a comprehensive understanding of the specific input variables and their real-world implications, the decision maker's grasp of the situation remains incomplete. SHAP also requires exponential computational time \cite{kjersti2019}.
Additionally, these methods are not very robust-- small changes in input features can lead to dramatic changes in explanations \cite{alvarez2018}.  
While useful for fine-tuning algorithms, these methods are not suited to real-time, high-stakes decision making tasks.

Given these limitations, there is a growing recognition that supporting the human's real time decision making process may require approaches that circumvent the need to understand the specific algorithms used by AI recommender systems.
To this end, one promising direction is to  provide contextual information about the task environment. Promising results have shown that this technique is effective in supporting the human's Orientation process when making a high-stakes decision \cite{srivastava2022, srivastava_lilly2024, srivastava2024}.

\subsection{Contextual Information Improving Team Performance with Recommender Systems}

To address the concerns, we conducted a series of experiments to provide an alternative to explaining or interpreting complex AI models by investigating the use of a model-agnostic method of supporting operators' cognitive process of judgement. We provided the decision maker with relevant information that the AI used to generate possible courses of action in an attempt to align the user and AI understanding of the state of the decision environment.  Our underlying hypothesis was that if the user and the AI had a joint understanding of the decision environment that a well-functioning AI would provide a recommendation that should be understandable by the user.

We began by, testing the basic hypothesis that providing the human decision maker with contextual information about the decision environment would align their situation awareness with that of a perfect AI recommender system \cite{srivastava2022}. Results showed that providing contextual information about the decision environment boosted the team's shared situation awareness. The relevant contextual information aided the human's judgement which whas shown by increased agreement between the human's final decision and the perfect AI's suggestion.

%Experiment 2: Supporting Human Judgement with Imperfect AI

Next we  investigated if supporting the human's judgement via contextual information would be effective with an imperfect AI\cite{srivastava_lilly2024}. The results of the second experiment indicate that supporting the human's judgement via contextual information helped participants determine the AI's error boundary (both when it is wrong and in what ways/situations it is wrong). It also reduced participants' overreliance on the AI's suggestion, as well as accurately calibrated the participant's trust in the AI's capabilities.

Encouraged by the findings from these initial experiments, the technique of supporting human situation awareness by providing them with relevant decision environment information seemed valid, but we wanted to verify that these findings are not due to the specific representations or visualizations of the contextual information shown to participants. Here the specific representations of the contextual information was varied to both increase and decrease the level of abstraction of the information.  Results from the next experiment \cite{srivastava2024} validated that crucial metrics in team decision-making, such as task performance, shared situation awareness, and trust in AI's capabilities, are robust to the level of abstraction at which information is displayed. This means that the supporting the human's judgement is important and helpful even if the information available to them is abstract or in overwhelming levels of detail.

Finally, in this work, we seek to compare the effectiveness of supporting human judgement via contextual information versus the popular transparency technique of providing local decision point explanations. This will inform whether or not this technique holds up to the current standard in automation transparency and support it as a valid transparency technique.

%JUST KIND OF ENDS HERE, NO GOOD WRAP UP OR LEAD INTO NEXT SECTION.

\section{Methodology}

We conducted a between-subjects experiment in which participants were asked to work with an automated recommender system to complete a decision making task.
We manipulated the type of transparency technique the participants were provided with, and we measured the effects on final task performance, trust in the recommender system, user workload, and the shared situation awareness between participants and the recommender.
There are four treatments groups: 1) Absent (the Baseline condition with no contextual information or explanation), 2) Contextual Information, 3) Local Explanation, and 4) Contextual Information AND Local Explanation.
Each treatment incorporated a different transparency structure intended to increase the human's understanding of the task and autonomous partner.
This experiment uses data from 180 participants. 
Data for the baseline (Absent, in which there is no explanation or contextual information) performance and the data for Contextual Information are from previous experiments, the results of which are published in \cite{srivastava2022, srivastava_lilly2024}. 

\begin{figure*}[!h]
     \begin{subfigure}{0.99\textwidth}
        \centering
        \includegraphics[width=\textwidth]{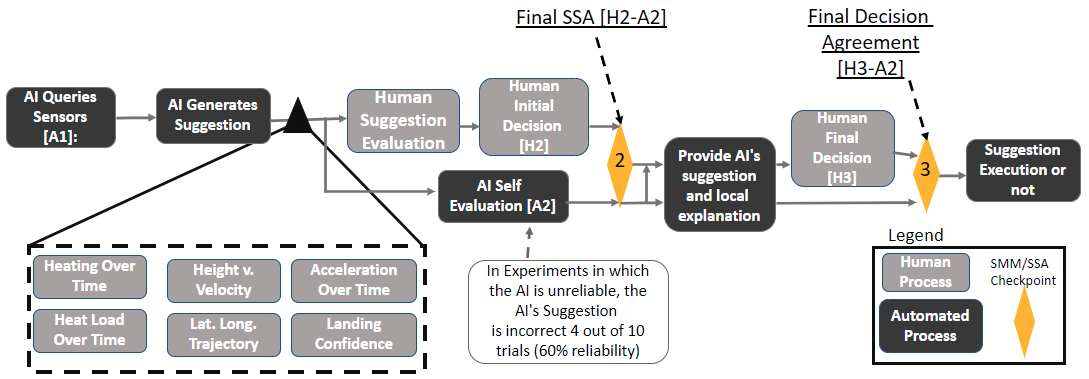}
    \end{subfigure}
    \caption{Task Outline and Metrics for EDL Trajectory Planning}\label{fig:exp5_task_outline}
\end{figure*}

For the last two conditions, we utilized the same scenarios with a different transparency method in which participants are assigned to either the third treatment group (Local Explanation) or the fourth treatment group (Contextual Information AND Local Explanation). 
Instead of or in addition to providing contextual information, we provide a local explanation (Fig. \ref{fig:local_exp}) at the time of the decision point (see Fig.\ref{fig:exp5_task_outline}). 
The explanation was a simple text box displaying a one-sentence explanation.  The explanation was developed using a the heuristic which explained why the AI made its decision.  
The explanation was always aligned with the AI’s (correct or incorrect) suggestion, such that, in situations in which the AI's decision was correct and corresponded to the ground truth, the explanation provided 2-3 corresponding figures of merit to cite as the reason why. Similarly, in the situations in which the AI's decision was incorrect, the explanation provided 2-3 contradicting or ambiguous figures of merit to cite.

We additionally include Trajectory Awareness as an independent variable that manipulates the amount of interaction the human has with the AI's generated solutions (Observation-Only or Interactive).

Participants were recruited through an online recruitment platform (Prolific, www.prolific.co).
Individuals who were under 18, located outside of the USA, not proficient in English, and/or did not have normal or corrected-to-normal vision were excluded from the studies.
%\textbf{@DIVYA INCLUDE EXCLUSION CRITERIA HERE}.
%The experiment to collect data for the latter two conditions was conducted online and collected data from 90 participants.

\subsection{Task Domain  \& Decision Environment} 
This work uses the same experiment platform and task developed by Srivastava et al \cite{srivastava2022,srivastava_lilly2024, srivastava2024}. This section describes the main components that are necessary to the development and understanding of this specific body of research work.
Participants played the role of commander of a spacecraft in Mars orbit, tasked with finalizing the \gls{EDL} trajectory of a probe to a landing site.
Participants were told that they were aided by an AI Mission Computer that made a parallel evaluation of the proposed trajectory and offered agreement or disagreement with the participant's decision.
After seeing the AI's recommendation, the participant had the final call on whether to execute or abort the mission.
We did not impose time constraints on the task.
For control and reproducibility, though the system was presented to participants as ``intelligent'', its responses, the contextual information it used, and the explanations it provided in each scenario were predetermined and fixed. 
The scenarios and tasks were also constant across participants, though the order in which they were shown was balanced.
These were all originally generated from a physics based simulation, where specific scenarios which possessed the correct mix of characteristics were chosen. 
The task is outlined in Fig.~\ref{fig:exp5_task_outline}.

%\begin{figure}[h]
%    \centering
    %\includegraphics[width=0.45\textwidth]{images/task_outline.png}
    %\caption{Task Outline for \gls{EDL} Trajectory Planning}
    %\label{fig:task}
%\end{figure}

\begin{figure}[h!]
    \centering
    \includegraphics[width=.45\textwidth]{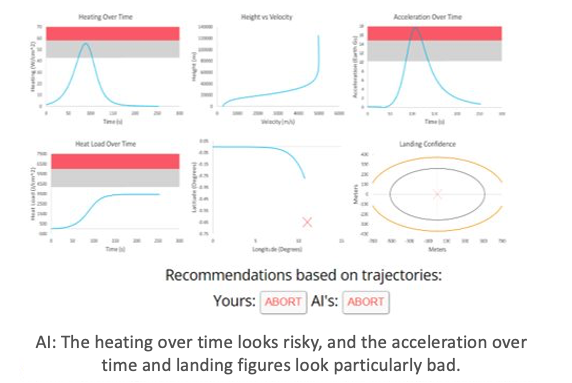}
    \caption{Example of AI's Local Explanation with the generated suggestion}
    \label{fig:local_exp}
\end{figure}

First, the AI Mission Computer generated a possible trajectory for the probe's landing (using algorithms unavailable to the human participant). 
The team was shown a set of six figures of merit (FoM) (Fig. \ref{fig:local_exp}) that characterized the proposed flight plan: velocity vs. altitude; heating rate, heat load, and acceleration vs. time; latitude vs. longitude; and landing confidence. 
The FoM charts were shaded to indicate safe, risky, and dangerous thresholds, and the participants were instructed on how to interpret each.
Using the FoMs, the participant and the AI individually evaluated whether or not to execute the landing trajectory (second judgement point represented by yellow diamond \#2). 
The AI Mission Computer's decision was then made known to the participant, along with an explanation as to why the AI made the suggestion it did in regard to that specific mission .

After seeing the explanation, the participant made the final decision on whether to execute or abort the mission in light of the AI's recommendation (third decision point represented by yellow diamond \#3).

\subsection{Experiment Design and Task Procedure}
%This section details the experiment design and the tasks associated with each phase of the experiments.
This between-subject experiment includes two independent variables:  Transparency Technique (4 levels) and Trajectory Awareness (2 levels). The ``AI'' behavior utilized in these studies is fixed and static to ensure that certain inputs (world state conditions) mapped to outputs (trajectory charts), thus allowing repeatability in the experiment.
To simplify the analysis, the decision to execute or abort is pre-determined for each scenario, allowing for repeatable manipulation of the accuracy of the AI's suggestion. For all of the data analyzed in this study, the AI is correct 60\% of the time, and fails in specific weather conditions. 
The decision to have the AI fail in a predictable way mimics real world situations in which algorithms that are trained on clearly labeled data and well-defined problems are successful when encountering similar situations, but fail when faced with noisy, degraded, or unfamiliar data. 
Additionally, by having the AI fail in a predictable manner, we were able to discern if providing transparency by way of environment information or by way of explanation helps the human understand the AI's limitations holistically (does the human pick up on the fact that the AI is giving inaccurate suggestions?) and specifically (under what circumstances does the AI fail?). 
These answers will inform the human's understanding of their AI partner, and thus their collaboration with it.  Trajanctory Awareness indicates whether or not the participants were required to classify each of the FoMs as good or bad (see Fig.~\ref{fig:local_exp}), or if they were simply presented with the FoMs with no requirement for classification.  This provided an additional control to understand if the participant adequately understood the FoMs and if this added task would improve performance.

Participants were assigned to a single treatment that was one of eight combinations of the Transparency Technique and Trajectory Awareness levels.  Participants completed 10 tasks within their treatment group. 

\begin{figure*}[]
     \begin{subfigure}{0.33\textwidth}
        \centering
        \includegraphics[width=\textwidth]{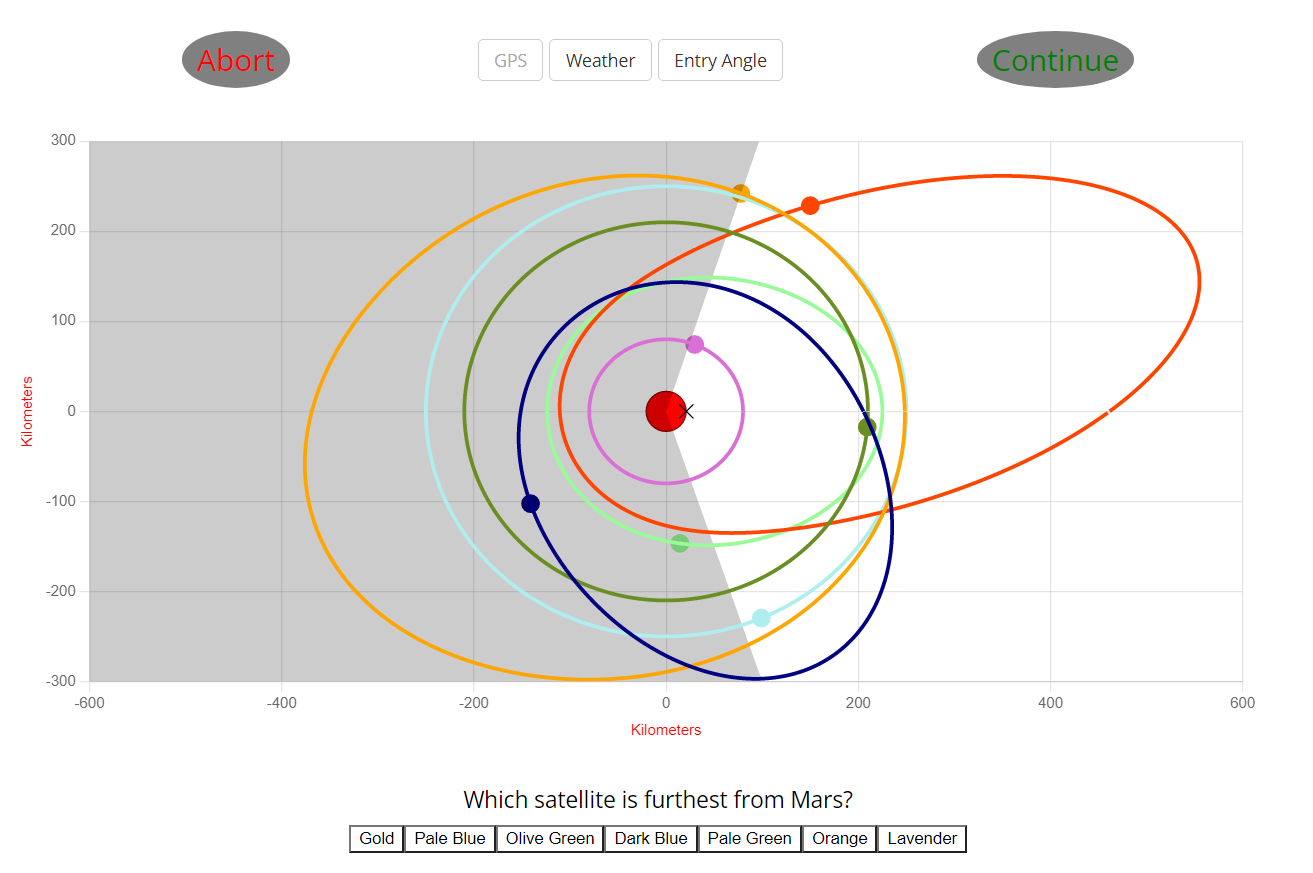}
        \label{fig:m1}
    \end{subfigure}
    \begin{subfigure}{0.33\textwidth}
        \centering
        \includegraphics[width=\textwidth]{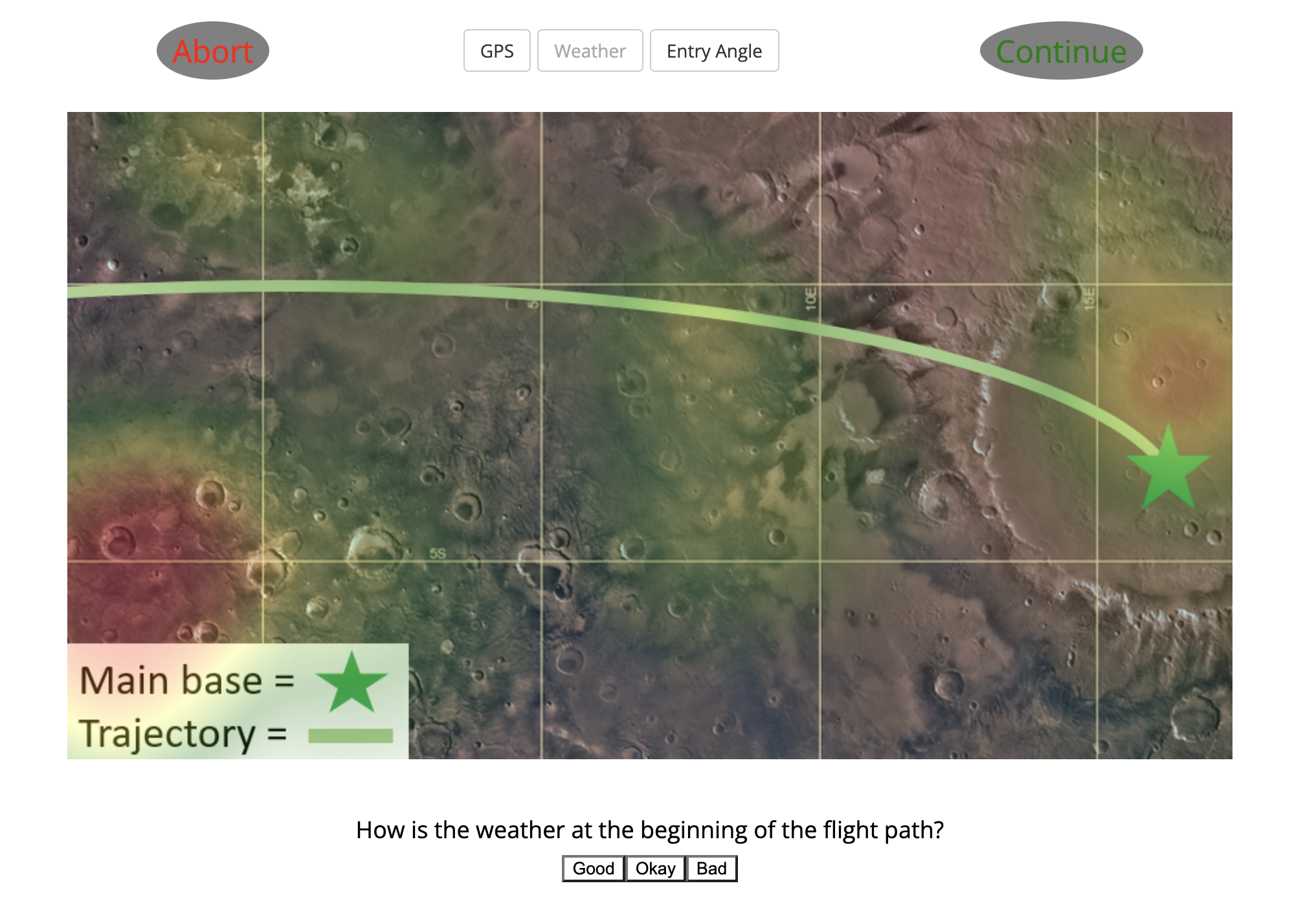}
        \label{fig:m2}
    \end{subfigure}
    \begin{subfigure}{0.33\textwidth}
        \centering
        \includegraphics[width=\textwidth]{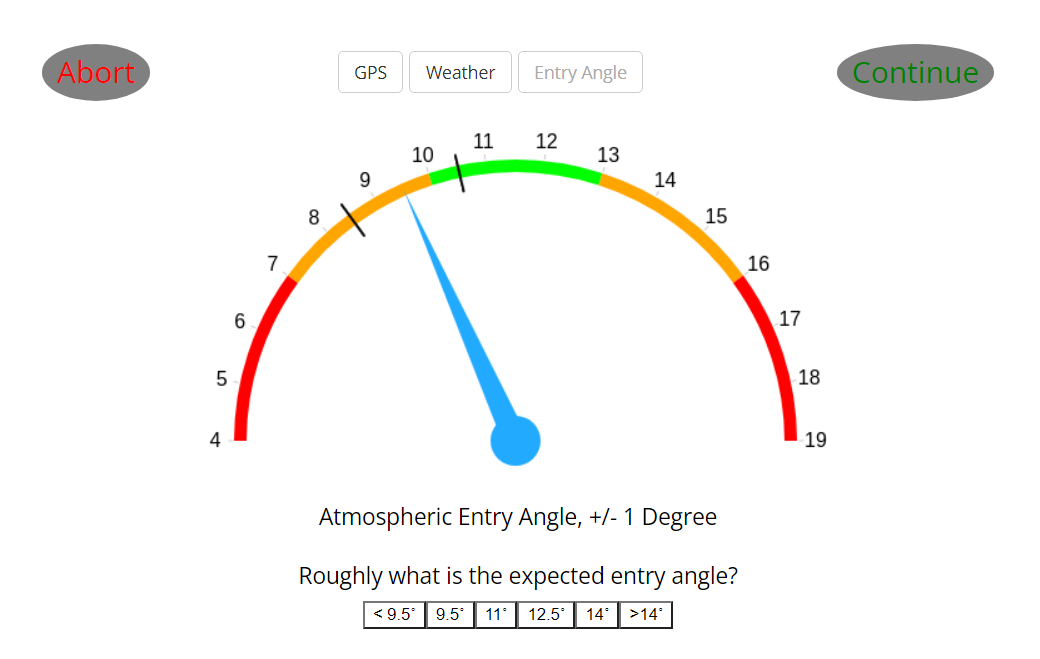}
        \label{fig:m3}
    \end{subfigure}

    \caption{World State Information from Left to Right: GPS, Atmosphere/Weather, Anticipated Entry Angle}\label{fig:ws_info}
\end{figure*}

The first transparency technique level is the Absent group which was the control group, in which participants were not given any relevant information about the world state.
%(World State Awareness = None in Table \ref{tb:TreatmentLevels2}).
This lack of contextual information mimics current black box systems, and participants in this group started directly at the Trajectory Evaluation part of the experiment.
In the Contextual Information mode, participants viewed three information screens about the world state (Fig. \ref{fig:ws_info}). The agents separately evaluated how risky the world state was (risky conditions vs safe conditions) in that scenario, and were asked if the world state conditions were risky or safe enough to attempt a landing.
The AI Mission Computer prepared its own answer to the same question.
For implementation purposes, for all phases, these responses were always correct.  This was the first assessment of the participant and AI having a shared understanding, i.e. the world state.  
The participant was then informed of the AI's judgment of the world state conditions (and thus whether they are in agreement).

%\subsubsection{Trajectory Evaluation}

 \begin{figure}[h]
    \centering
    %\caption{AI-generated trajectory with interactive questions}
    \includegraphics[width=0.45\textwidth]{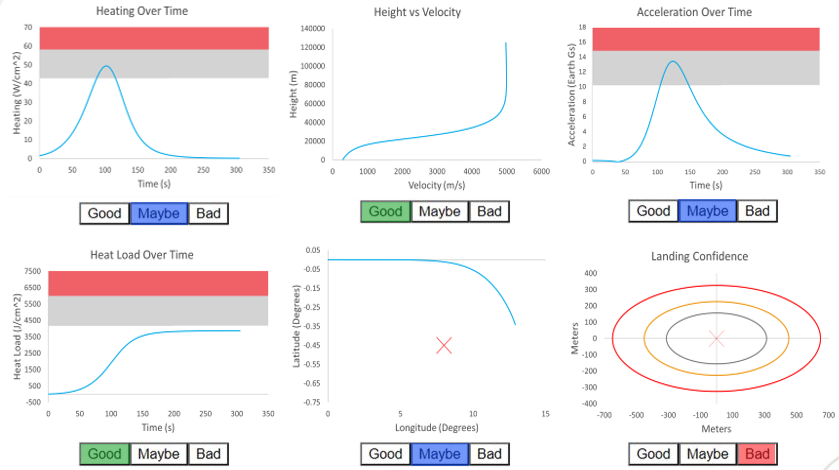}
    \caption{AI-generated trajectory with interactive questions}
    \label{fig:traj}
\end{figure}

Following the assessment of the world state context information to align situation awareness, a landing trajectory was presented, and the agents (both human and AI) evaluated whether or not to execute it.
For the Absent Group, and the Local Explanation Group, the experiment started here as there was no contextual information to review. 
Participants were shown a set of six figures of merit that described the proposed flight plan.
The Trajectory Awareness variable had two levels: Observation-Only or Interactive.
In the Observation-Only Mode, participants were able to view the six figures of merit for the suggested trajectory and made an execute/abort decision.
In Interactive Mode, participants were additionally asked to mark each of the six figures of merit as ``Good", ``Bad", or ``Maybe" according to that chart's specific risk factors (Fig. \ref{fig:traj}).
After observing or interacting with each of the trajectory charts, participants were asked to decide whether to execute the trajectory.  
This decision point allowed us to get a second measure of shared understanding between the agents.  

The AI Mission Computer's decision was revealed to the participant after their own decision was entered. For the groups that included a Local Explanation, the decision came along with an explanation that described why the AI made the suggestion it did (Fig. \ref{fig:local_exp}).

Participants were then asked to make a final decision,  allowing us to understand how much the AI's own assessment influenced the participant.

To summarize, a total of 2 decision points were recorded for the participants in the Absent and Local Explanation groups (their initial decision before seeing the AI's recommendation, and then final decision). A total of 3 decision points were recorded for the participants in the  Contextual Information, and the Contextual Information AND Local Explanation groups, as they also had to make a judgement about the environment before moving on to the actual task decision. 
These decisions could be compared in several ways providing us with insights into the alignment between human and AI, consistency of participant decisions, the influence of the AI's analysis , the influence of the contextual information and the influence of the explanation on the participant.

\subsection{Measures and Dependent Variables}
From the participant decisions recorded, we computed the following metrics of interest:

\textbf{Final Agreement: }We first recorded the Final Agreement between the participant and the AI after the AI's recommendation to execute/abort was revealed (yellow diamond 3 Fig~\ref{fig:exp5_task_outline}, as at this point all participants have experienced some transparency technique.  We computed the Final Agreement on a per participant basis across all 10 scenarios they experienced.

\textbf{Task Performance: }We assessed the final decision of the human participant against the ground truth. This was a binary metric of whether the human made the right decision on whether to abort or execute the trajectory.

\textbf{Sway, the AI's Influence on the Participant: }  We recorded the number of decisions per participant that changed from initial (yellow diamond 1) to final decisions (yellow diamond 3) and from there computed the percentage of initial disagreements that were resolved.  As the AI did not change its decision, and the situations in which the AI made decisions that were in error were known, this allowed us to understand the appropriate or detrimental influence that the AI had on the participant.

\textbf{Time: }  We recorded the completion time for participants to complete the mission planning task.  This metric allowed us to understand the additional burden that contextual information and explanations would add to the baseline task.  Additional time is important as often contextual information is excluded as it is considered too temporally costly or mentally taxing to be included.

Additionally, responses were recorded from the  subjective questionnaires given to participants.

\textbf{Trust: }The pre-experiment questionnaire was an i-THAu trust assessment \cite{ithau} in which participants were asked to answer a series of statements about their Faith in Persons and Faith in Technology on a seven-point Likert scale from [-3:3].
A composite average of their answers informs their overall dispositional trust in these two categories.  Following the experiment the rest of the the i-THAu trust assessment in which participants responded to a series of statements about their experience of working with the automated system on a seven-point Likert scale from [-3:3] was administered.
For both i-THAu assessments, a rating of -3 indicated a lack of trust -- the subject didn't trust people/technology or depend on the AI to help them with the \gls{EDL} task, or they didn't understand the role of the AI.
Conversely, a rating of 3 indicated a high level of trust in people/technology, as well as high trust in and understanding of the AI.

\textbf{Workload: }
We measured participant workload using the NASA TLX workload assessment which measured overall cognitive workload.

\subsection{Experiment Procedure}
Participants were given briefing documents and asked to complete a consent form upon starting the study. Each study had four main components: pre-experiment questionnaires, training session, data collection, and post-experiment questionnaires. 
The pre-experiment questionnaires measured dispositional trust of the participant in two areas: Faith in Technology and Faith in Persons. These two factors were measured to see if they could predict the accuracy of the human’s mental model and overall task performance. 
Participants were assigned randomly to a treatment group, and the order of scenarios shown to participants was balanced. Instructional videos, tailored to the specific treatment, were used to train participants on the task and the AI Mission Computer. 
No task-relevant experience was assumed. 
Participants then completed 6 practice rounds of the mission planning task. 
Participants who had correct task performance in 5 out of 6 practice trials were allowed to proceed past the practice rounds. 
Passing participants then proceeded to complete 10 trials of the mission planning task. 
Participants were given feedback on whether their decision was correct or not during the practice rounds, but they were not given any feedback on their performance during the actual data collection rounds. 
%At the end of the experiment, participants completed two final questionnaires: 1) NASA TLX \cite{hart2006} and 2) i-THAU trust assessment . 
The post-experiment questionnaires measured workload and learned trust. %Cognitive workload frustration, and perceived performance spoke to the human’s experience in completing the AI-advised decision-making task. Situational and learned trust spoke to the human’s impression and mental model of the AI.

\section{Results}
For all objective metrics, we performed a linear mixed effects analysis to determine the significance of the relationship between the metric and our independent variables.
The fixed effects were the Transparency Technique (Absent, Contextual Information,  Explanation, Contextual Information and Explanation), Trajectory Awareness (observation or interaction), the average Dispositional Trust in People and the average Dispositional Trust in Technology.
We also included an interaction effect between Transparency Technique and Trajectory Awareness and intercepts for participants as a random effect.
We performed an ANOVA for each fitted linear mixed-effects model.

\subsection{Objective Metrics}

\subsubsection{Final Decision Agreement}

\begin{figure}[h!]
    \centering
    \includegraphics[width=0.45\textwidth]{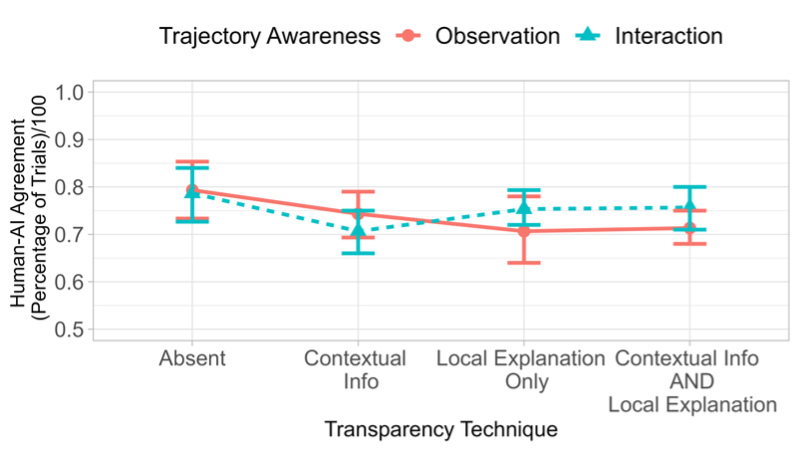}
    \caption{Final Agreement [\%] Between Human and 60\% Accurate AI}
    \label{fig:exp4_final_agr}
\end{figure}

The team’s Final Decision Agreement (after the AI reveals its decision, shown as yellow diamond \#3 in Fig. \ref{fig:exp5_task_outline}) is shown in Fig. \ref{fig:exp4_final_agr}. All transparency techniques (providing contextual information, a local explanation, or both) result in a team agreement trending closer to the ideal 60\% agreement than having no alignment or explanation technique (Absent). However, a linear mixed effects analysis (Table \ref{tab:final_agr_anova_exp4}) for this metric revealed that none of the fixed effects were statistically significant in predicting the Final Agreement between the team.

\begin{table}[h!]
\centering
\caption{ANOVA for Final Decision Agreement}%
\begin{tabular}{@{}llll@{}}
\toprule
 & df, Error & F & P \\
 \hline
Transparency Technique & 3, 172 & 2.024 & 0.1123  \\
Trajectory Awareness & 1, 172 & 0.232 & 0.6305   \\
Transp. Techique x Trajectory Aw. & 3, 172 & 1.337 & 0.2639 \\
\hline
\end{tabular}
\label{tab:final_agr_anova_exp4}
\end{table}

\begin{figure}[h!]
    \centering
    \includegraphics[width=0.45\textwidth]{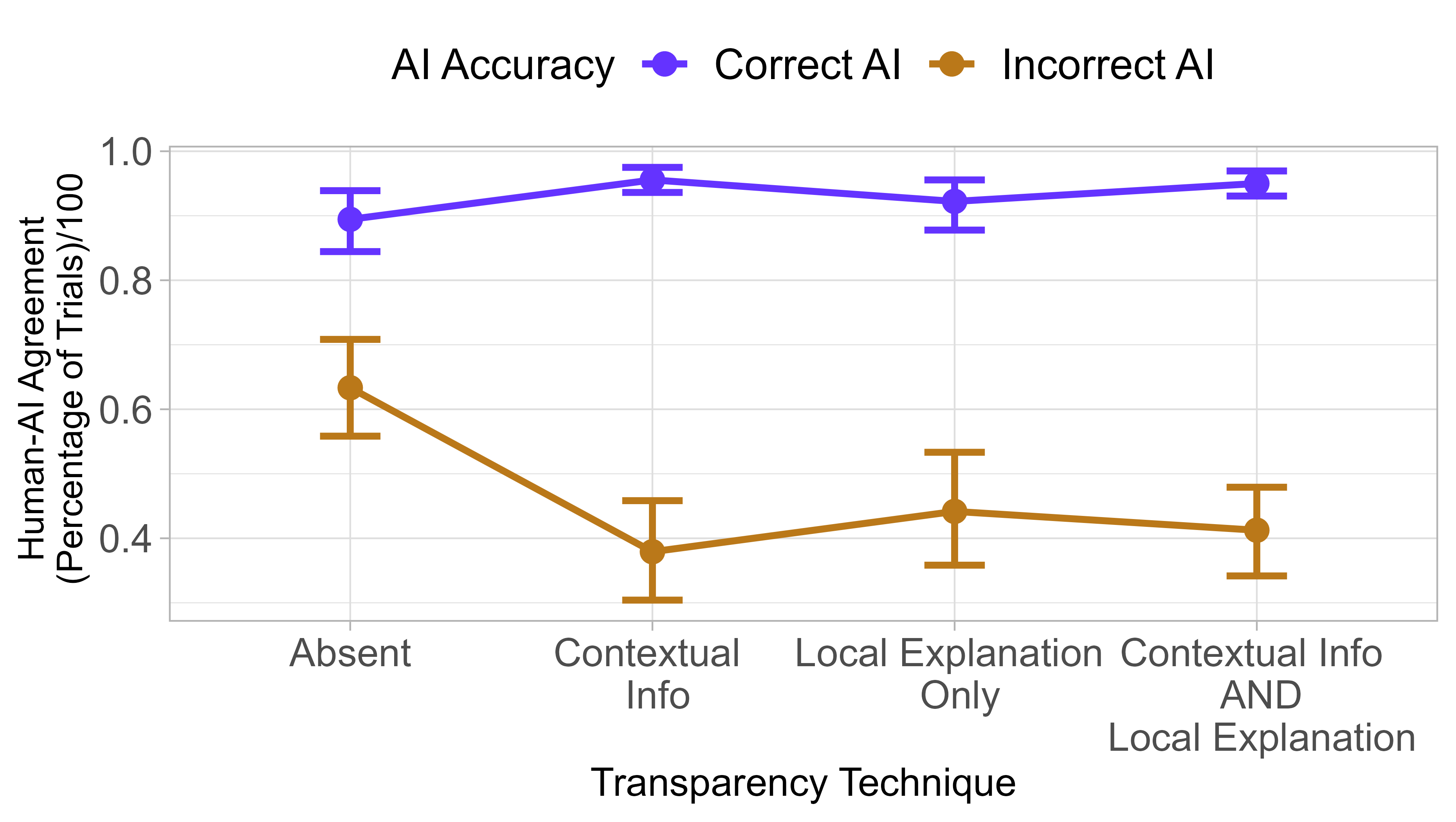}
    \caption{Final Decision Agreement [\%] Between Human and 60\% Accurate AI for both Trajectory Awareness Levels}
    \label{fig:exp4_final_agr_acc}
\end{figure}

Separating out the influence of the AI reliability, Figure \ref{fig:exp4_final_agr_acc} depicts how the trends for the team’s Final Decision Agreement were influenced by the reliability of the AI. In this figure, we collapsed across trajectory awareness for clarity as it had not previously shown to influence this metric. 
Ideally, when the AI is correct, the human would agree with it 100\% of the time; conversely when the AI is incorrect, that agreement would drop to 0\%. 
The average final agreement between the participant and the AI when the AI was correct (purple line in Fig. \ref{fig:exp4_final_agr_acc}) trended upward toward 100\% when the participant worked with an AI that was transparent in some way. 
We also see that the average Final Agreement between the participant and the AI when the AI was incorrect (brown line in Fig. \ref{fig:exp4_final_agr_acc}) trended downward toward 40\% if the participant was aided by a technique to understand the AI Mission Computer. 
This indicates that providing contextual information or a local explanation to the human improves their mental model of their AI partner (understanding that their AI teammate is limited in some capacity). 
In turn, this improvement in the human’s mental model helps them determine when (and when not) to align with the AI’s suggestion for the final decision. 
Further this indicates a bias toward AI agreement, particularly in the incorrect cases, that can be reduced, but not eliminated, with the introduction of transparency techniques.

%\kf{Did you run a version of the model with the AI Accuracy as a variable ?  Possibly even splitting the data set to only look at the Incorrect AI runs? }

\subsubsection{Task Performance}

\begin{figure}[h!]
    \centering
    \includegraphics[width=0.45\textwidth]{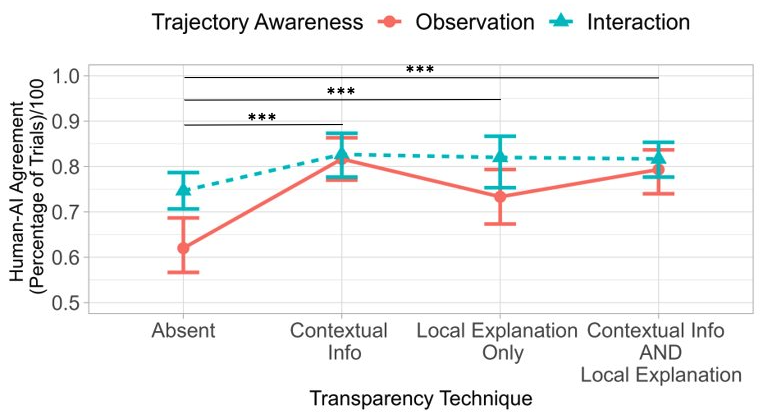}
    \caption{Percentage of trials in which the Human's Final Decision is Correct (Task Performance)}
    \label{fig:exp4_tp}
\end{figure}

Table \ref{fig:exp4_tp} shows the percentage of trials in which the human’s final decision was correct (Task Performance). The Absent level of working with the black box AI results in an average of ~70\% Task Performance, whereas having some transparency technique increased task performance to an average of ~80\%. A linear mixed effects analysis revealed a statistically significant difference between the baseline and each of the transparency techniques, but not between techniques. This means that having at least some sort of decision-making support yields higher team performance, as expected. Additionally, it was noted that explanation performed similarly to contextual information.

\begin{table}[h!]
\centering
\caption{ ANOVA for Task Performance}%
\begin{tabular}{@{}llll@{}}
\toprule
 & df, error & F & P \\
\hline
Transparency Technique & 3, 172 & 8.787 & $<$0.0001  \\
Trajectory Awareness & 1, 172 & 	6.199 & 	0.0137   \\
Transp. Techique x Trajectory Aw. & 3, 172 & 1.862 & 	0.1378 \\
\hline
\end{tabular}
\label{tab:tp_exp4}
\end{table}

The results from the ANOVA for this metric (Table \ref{tab:tp_exp4}) indicate that access to a transparency technique is statistically significant in influencing how correct the participant's final decision is. No other effects were significant. 

\subsubsection{Sway, the AI's Influence on the Participant}

%*** Decide if characterizing as resolved team disagreements or gap in human and ai's final "judgement" (decision? establish consistent terminology) ***  KF-if you do this you need to put it up in the metric section

\begin{figure}[h]
    \centering
    \includegraphics[width=0.45\textwidth]{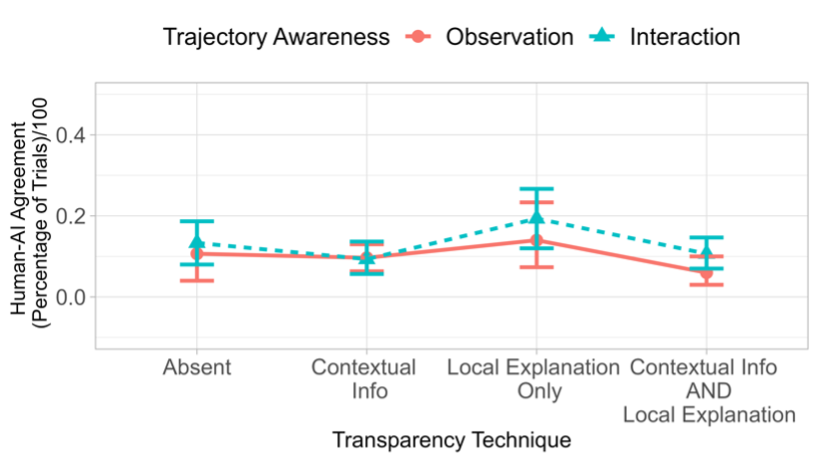}
    \caption{Gap in Human's and AI's Final Judgement}
    \label{fig:exp4_sway}
\end{figure}

Figure \ref{fig:exp4_sway} shows the gap in the participant’s and AI’s final judgement, operationalized as the percentage of decisions per participant that changed from initial to final decisions to match the AI’s suggestion (positive being changes toward the AI stance). 
Providing just the local explanation resulted in slightly higher deference to the AI’s suggestion than providing only or additional contextual information. 
This indicates that, while having the local explanation yields equivalent Task Performance as having contextual information, it increases reliance on the AI partner, whereas having contextual information supports the human decision maker’s judgment to be on aligned with the AI partner when appropriate and to be informed sufficiently to contribute to a joint decision.

\begin{table}[h!]
\centering
\caption{ ANOVA for Resolved Team Disagreements}\label{tab:sway_anova_exp4}%
\begin{tabular}{@{}llll@{}}
\hline
 & df, error & F & P \\
\hline
Transparency Technique & 3, 172 & 3.469 & $<$0.0001  \\
Trajectory Awareness & 1, 172 & 2.345 & 	0.1275   \\
Transp. Technique x Trajectory Aw. & 3, 172 & 0.558 & 0.6436 \\
\hline
\end{tabular}
\end{table}

The results from the ANOVA for this metric (Table \ref{tab:sway_anova_exp4}) indicate that having access to a transparency technique is statistically significant in predicting when the participant will change their decision to align with the AI’s suggestion. 

\subsubsection{Average Completion Time}

\begin{figure}[h]
    \centering
    \includegraphics[width=0.45\textwidth]{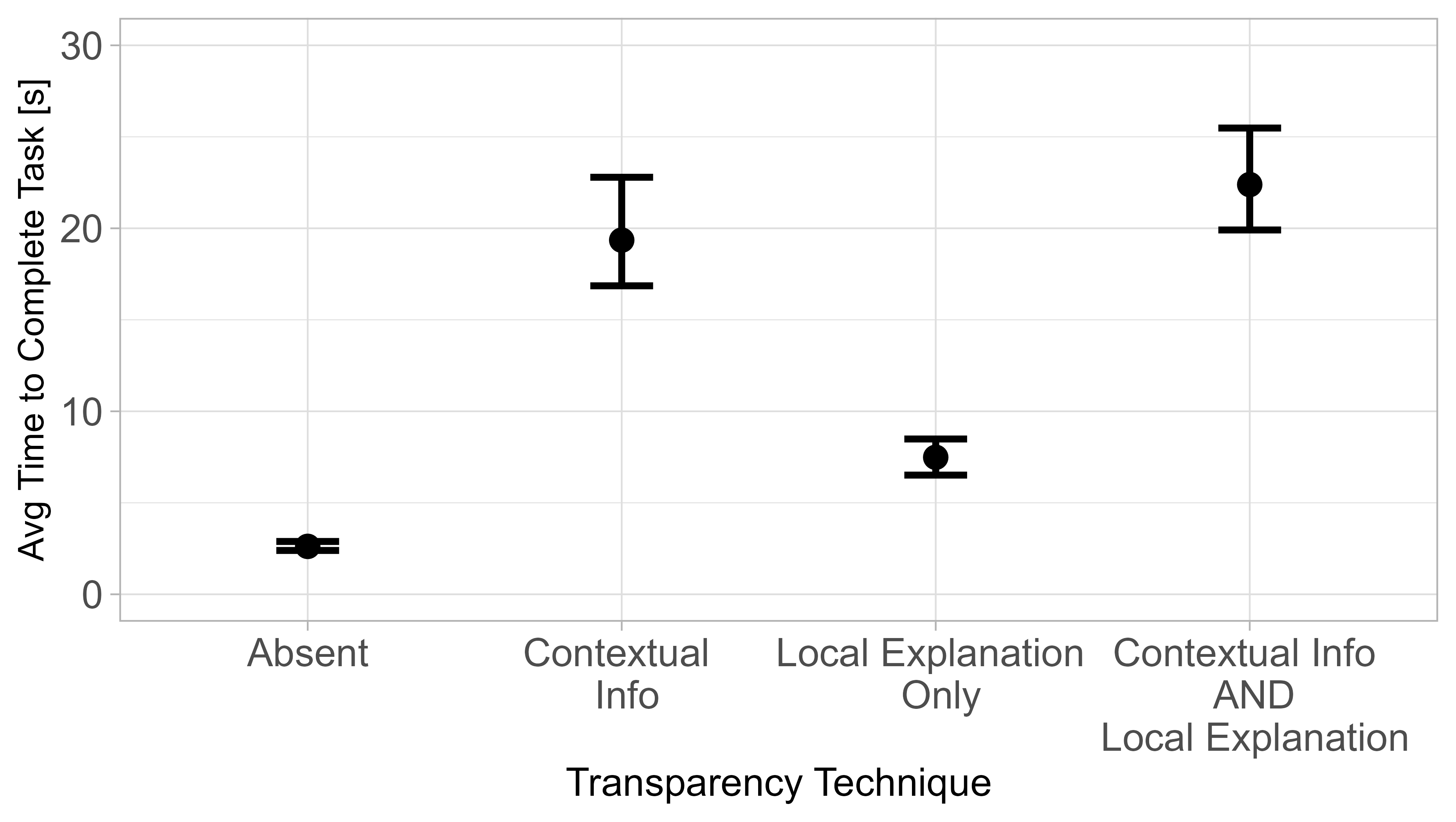}
    \caption{Average Time to Complete a Mission-Planning Trial}
    \label{fig:exp4_time}
\end{figure}

\begin{figure}[h]
    \centering
    \includegraphics[width=0.45\textwidth]{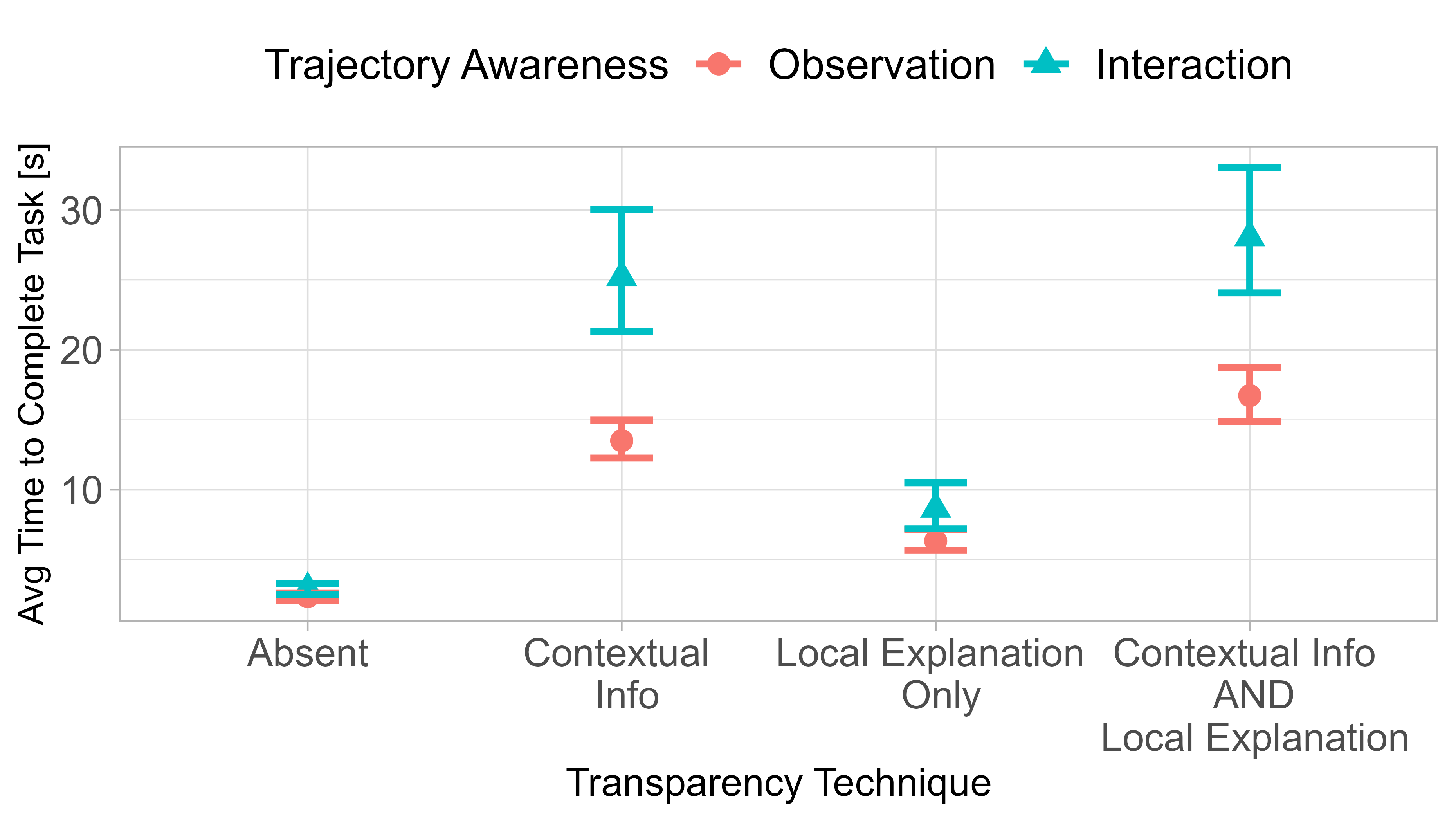}
    \caption{Average Time to Complete a Mission-Planning Trial}
    \label{fig:exp4_time_traj}
\end{figure}

Finally, we observed the additional objective metric of time. Figure \ref{fig:exp4_time} shows the average time it took for a participant to complete the mission planning task. 
In our control group, where participants had no additional information regarding the mission planning task or their AI partner, participants took roughly 3 seconds to complete the task. 
Subsequently, adding more transparency to the system, and thus more for the decision maker to contend with, resulted in higher completion times. 
However, we see from Fig. \ref{fig:exp4_time_traj} that interacting with the figures of merit was the real driver of additional completion time.  Individually assessing each FoM had a significant impact on the average completion time for those who saw contextual information as well.

\subsection{Subjective Metrics}

\subsubsection{NASA TLX}

\begin{figure*}[h]
    \centering
    \includegraphics[width=\textwidth]{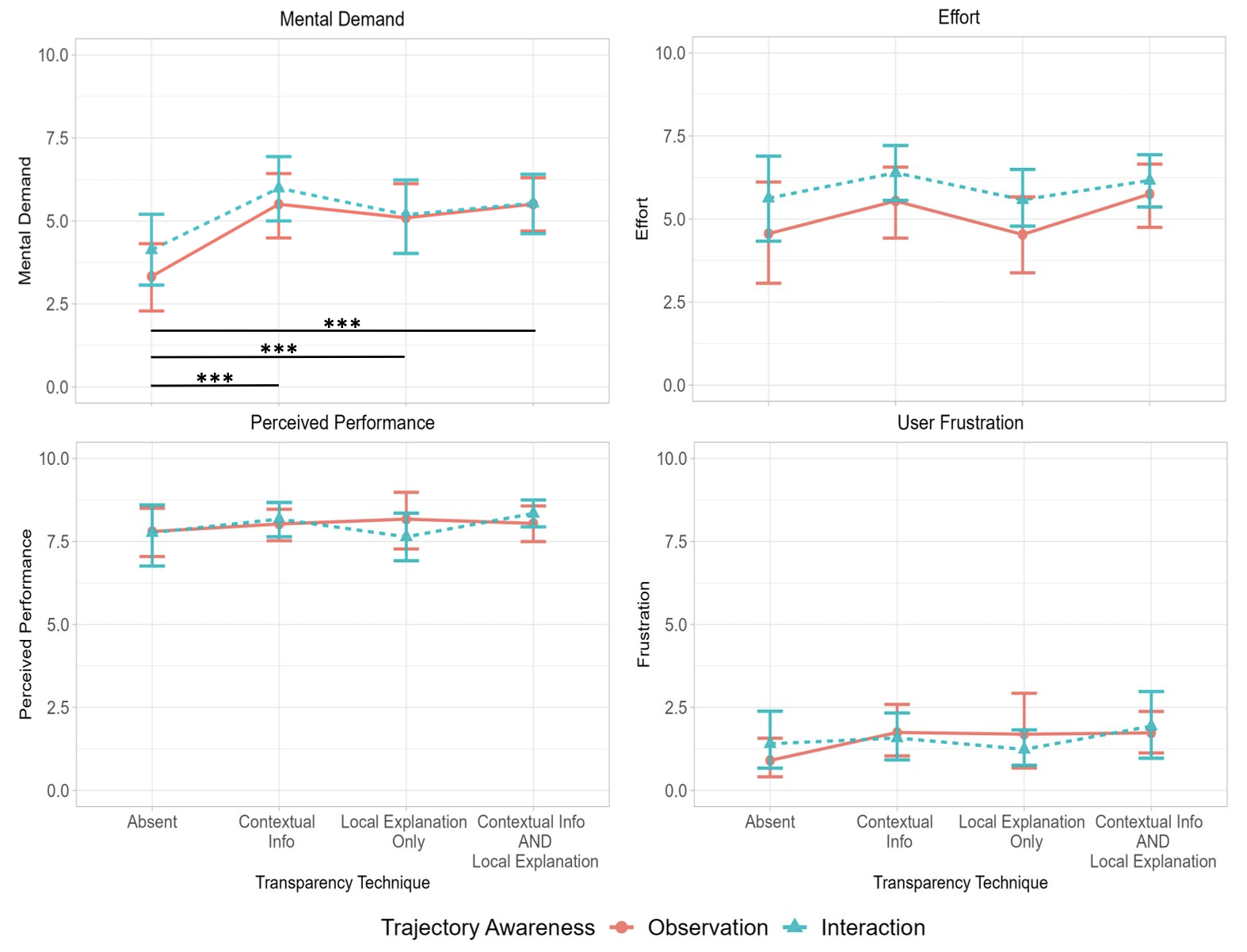}
    \caption{Composite Results of NASA TLX Questionnaire}
    \label{fig:exp4_tlx}
\end{figure*}

Figure \ref{fig:exp4_tlx} shows the subjective mental workload of each participant using the NASA TLX assessment at the end of the experiment. While all six subscales were recorded as there was little physical activity and no temporal constraints we have chosen only to present the remaining four subscales.  Having any transparency technique yields more effort, mental demand, and frustration on the user’s part, it also yields a slightly higher perception of their own performance. 
Only Mental demand indicated any significant impact from the Transparency Techniques and only when compared to the baseline of nothing.
Additionally, while contextual information generated slightly more mental demand and effort than providing only a local explanation, both techniques elicited the same amount of frustration and perceived performance, while contextual information also bridged the gap between the human and AI’s judgement (Fig. \ref{fig:exp4_sway}).

\subsubsection{i-THAu}

\begin{figure}[h!]
    \centering
    \includegraphics[width=0.45\textwidth]{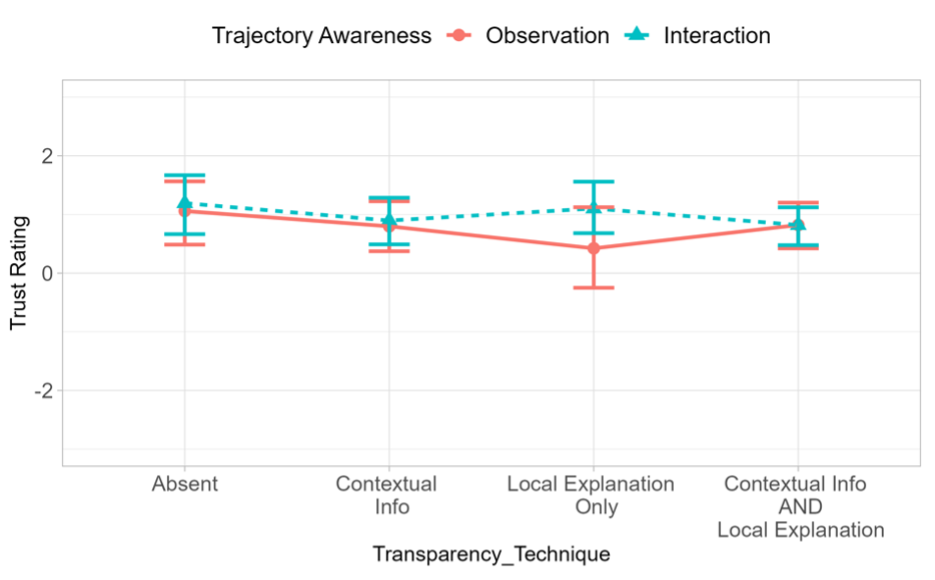}
    \caption{Post-Experiment i-THAu Metric of Human's Trust in AI's Capabilities}
    \label{fig:exp4_trust}
\end{figure}

Figure \ref{fig:exp4_trust} shows the human’s trust in the AI’s capabilities to complete the task measured in the post-experiment i-THAu assessment. 
Across all treatments trust remained in the low positive range with scores close to 1 on a scale from -3 to 3.  
Participants who worked with a black box AI (Transparency Technique = Absent) tended to have a slightly higher trust in the AI’s capabilities than participants who had more insight into the AI’s decision making. 
However, this metric is not statistically significant for differences in trust in capability between transparency techniques. 
Fig. \ref{fig:exp4_trust} also indicates that having some sort of guided interaction/information enforcement in the form of interaction increases trust in the AI when given a local explanation only. 
This may be because participants who only see a local explanation tend to defer to the AI’s suggestion more than those who only or additionally see contextual information (Fig. \ref{fig:exp4_sway}), and interacting with relevant questions forces participants to absorb domain information more, therefore increasing their understanding of the AI’s suggestion.

\section{Discussion}
The purpose of this series of experiments was to empower and most effectively utilize human decision makers when working with AI recommender systems. 
Rather than attempting to explain or interpret the model-specific workings of complex AI models, we examined an alternative approach designed to support the human decision-maker's cognitive process of judgment, i.e. assessment of the state of the world, by providing a model-agnostic alternative method. We evaluated equipping the decision-maker with the relevant information that the AI system utilized to generate potential courses of action. Our hypothesis was that this technique would align the particpant's situational awareness with that of the world in which the AI is operating. 
By establishing a shared understanding of the underlying information, e.g. shared situation awareness, we expected that the overall performance of the human-AI team would improve as the human would be grounded sufficiently to make an informed and independent decision.

This experiment was the last in a series of four that systematically evaluated the impact of contextual information to improve human judgment and align the human-AI situation awareness.  %The first experiment \cite{srivastava2022} aimed to determine if providing contextual information about the decision environment could align the human decision-maker's situational awareness with that of a perfect AI recommender system. The results showed that sharing this contextual information boosted the team's shared situational awareness. The world state information aided the human's judgment, as evidenced by the agreement between the human's choice to abort or execute and the perfect AI's suggestion. 

%The next experiment investigated whether supporting the human's judgment via contextual information would be effective with an imperfect AI \cite{srivastava_lilly2024}. The results indicated that providing contextual information helped participants determine the AI's error boundaries (when it was wrong and in what situations it was wrong). It also reduced participants' over-reliance on the AI's suggestions and accurately calibrated their trust in the AI's capabilities. 

%To validate the findings of the previous experiments and ensure they were not due to the specific visualizations used, the subsequent follow-up study \cite{srivastava2024} varied the levels of abstraction at which the contextual information was displayed to participants. The results indicated that crucial metrics in team decision-making, such as task performance, shared situational awareness, and trust in AI's capabilities, were robust to the level of abstraction at which information was displayed. This means that supporting the human's judgment is important and helpful, even if the available information is abstract or presented in overwhelming detail.

%Finally, with this last experiment,
Here we sought to compare the effectiveness of supporting human judgement via contextual information versus the popular transparency technique of providing local decision point explanations. We found that providing contextual information was a viable alternate solution to providing explanations, as overall performance was very similar. This means that, based on the context of the problem space, this novel approach may be less costly to implement and yield as effective results as understanding the AI's inner algorithms would.

This work has some limitations.
%...\kf{add some limitations here.}
This body of research work was conducted in a single domain, potentially restricting the broader applicability of its conclusions in other fields.
Nonetheless, we believe that supporting human judgment in decision-making scenarios where operators bear responsibility for their choices, especially in high-stakes cases, will likely serve to bolster performance across various fields. 

Furthermore, this study did not evaluate the efficacy of providing contextual information  against other methods of enhancing human judgment. Moreover, the AI mission planner employed in our experiments was a basic heuristics-based model, and we did not explore alternative AI explanation techniques beyond local explanations. 
Subsequent studies could investigate the effect different representations of contextual information could have on the human's judgment, or compare this technique to other XAI techniques.

\section{Conclusion}
These subjective metrics in conjunction with the objective metrics indicate that having at least some sort of transparency technique yields relatively equivalent, improved team performance with respect to having no transparency technique at all (Figs. \ref{fig:exp4_final_agr} and \ref{fig:exp4_tp}). 
However, having contextual information is more effective in reducing overreliance on AI (Fig. \ref{fig:exp4_sway}) and determining in what ways the AI is limited/where its error boundary is. 
Both transparency techniques were also effective in boosting the human decision maker’s perception of their own performance while negligibly increasing participant frustration. 
Additionally, we see that, although those who see world state information take more time to complete the task, they achieve significantly better objective performance than those who have no insight into the system. These findings are important because they emphasize the need for transparency in AI systems to achieve improved team performance, and this technique enables various solutions for increasing transparency for human-autonomy teams that can be determined based on system design and environment criteria.

\section*{Acknowledgment}
This work was funded by Sandia National Laboratory (SNL) with Dr. Paul Schutte serving as Program Manager.  
This work is solely that of the authors and does not represent an official SNL position.

\ifCLASSOPTIONcaptionsoff
  \newpage
\fi



%\bibliography{bibtex/bib/main.bib}{}


\bibliographystyle{IEEEtran}

\end{document}